\documentclass{IOS-Book-Article}

\usepackage{mathptmx}
\usepackage{soul}\setuldepth{article}
\usepackage{graphicx}
\def\hb{\hbox to 11.5 cm{}}

\begin{document}

\pagestyle{headings}
\def\thepage{}
\begin{frontmatter}              

\title{Trust in foundation models and GenAI:\\A geographic perspective}

\markboth{}{}

\author[A]{\fnms{Grant} \snm{McKenzie}%
\thanks{Corresponding Author: Grant McKenzie, grant.mckenzie@mcgill.ca}},
\author[B]{\fnms{Krzysztof} \snm{Janowicz}}
and
\author[C,D]{\fnms{Carsten} \snm{Ke{\ss}ler}}

\runningauthor{G. McKenzie et al.}
\address[A]{McGill University, Canada}
\address[B]{University of Vienna, Austria}
\address[C]{Bochum University of Applied Sciences, Germany}
\address[D]{Aalborg University Copenhagen, Denmark}

\begin{abstract}
Large-scale pre-trained machine learning models have reshaped our understanding of artificial intelligence across numerous domains, including our own field of geography. As with any new technology, trust has taken on an important role in this discussion. In this chapter, we examine the multifaceted concept of trust in foundation models, particularly within a geographic context. As reliance on these models increases and they become relied upon for critical decision-making, trust, while essential, has become a fractured concept. Here we categorize trust into three types: epistemic trust in the training data, operational trust in the model’s functionality, and interpersonal trust in the model developers. Each type of trust brings with it unique implications for geographic applications. Topics such as cultural context, data heterogeneity, and spatial relationships are fundamental to the spatial sciences and play an important role in developing trust. The chapter continues with a discussion of the challenges posed by different forms of biases, the importance of transparency and explainability, and ethical responsibilities in model development.  Finally, the novel perspective of geographic information scientists is emphasized with a call for further transparency, bias mitigation, and regionally-informed policies.  Simply put, this chapter aims to provide a conceptual starting point for researchers, practitioners, and policy-makers to better understand trust in (generative) GeoAI. 
\end{abstract}

\begin{keyword}
trust\sep foundation model\sep geoai\sep credibility
\end{keyword}
\end{frontmatter}
\markboth{May 2025\hb}{May 2025\hb}

\section{Introduction}

As Maria boards her flight bound for Vienna, she considers the complexity built into her travel.  The various components that contribute to her trip are based on algorithms that, while increasingly powerful and accurate, are also opaque to the end user.  The path of her flight has been pre-computed to take into account wind speed models, conflict area avoidance, costs, and alignment with multiple other flights operating in the same region at the same time.  The pilots flying the plane have gone through hours of flying simulations, all developed through the use of various machine learning models that read in variables ranging from posture to response time through eye-tracking.  Hiring for the cabin staff was done by first having an algorithm sort applications based on thematic topics and perceived fit within the company.  The exact meal portions, ingredients, and cooking temperatures were all determined based on analysis of existing data to optimize meal service and minimize food costs. Finally, her ticket price was determined in near-real time adjusting to her own context as well as the expected demand by others.

More and more of the components that go into a simple trip from Montr\'eal to Vienna are shaped by what is typically referred to as artificial intelligence (AI) models. Interestingly, most of the literature does not distinguish between AI and machine learning, although they are far from being the same. While machine learning models are trained to perform a specific task, such as classifying images according to the presence of cars or detecting text in images, AI systems generally mimic human behavior in some ways, such as chat bots used for customer assistance. Increasingly, the backbone of many of these AI systems are foundational models --  large-scale machine learning models, pre-trained on vast data sets, that can be fine-tuned for various downstream tasks and applications across different domains.  Put differently, an AI chatbot such as ChatGPT is not just the underlying language model (GPT). As will be discussed later on, trusting an (open-weights) model is different from trusting the bot developed on top. 

While we have seen dramatic improvements in these systems in recent years, the complexity of these models and deployments has also led them to be less transparent.  We are no longer able to understand decisions made by discussing with the decision maker or investigating the models ourselves, and, increasingly, we are forced to ``trust'' AI outputs.  In most cases, this means trusting the models, trusting the data they have been trained on, trusting the model developers, trusting prompt engineers, alignment teams, secure web infrastructure, and data/model curators, and trusting the results, recommendations, and policy decisions produced as outputs of these models/systems.  For a technology that is in the midst of disrupting the lives of virtually every person on our planet, trust appears to be at the center of the entire process. Yet, we are facing a sudden political/cultural shift, where oversight over technologies that can potentially rupture the very fabric of society is being scaled back.

\subsection{Foundation models}
While foundation models are one of the central themes of this book, it is important to first define what these models are and briefly describe their role and impact on geography as a discipline. Foundation models refer to large-scale, pre-trained models that are the basis (hence \textit{foundation}) for a wide variety of downstream tasks and applications. These models are often trained on massive heterogeneous data sets and are designed to be general-purpose, meaning they can be adapted or fine-tuned for specific (\textit{downstream}) tasks with relatively little additional training.  With the growing impact of these models, discussions about their role in shaping our world, e.g., our interaction with machines but also with other humans, are coming to the societal forefront.  With respect to the discipline of geography, or the geographic sciences, foundation models are ingesting massive amounts of ``geospatial'' data but also using this information to make geographically-informed decisions, generate recommendations and predictions based on geographic phenomena, and are beginning to generate geographically (and culturally) valid data~\cite{torrens2018artificial,zhao2021deep}.  Throughout this book, there are numerous examples of the role that geography plays in the development and training of foundation models, as well as the ways that foundation models shape the field of geography. While research on understanding and potentially mitigating the resulting biases (e.g., from training data, model parameters, etc.) has been gaining attention for a few years \cite{mehrabi2021survey,janowicz20233,liu2025operationalizing}, work on trust in GeoAI is lagging behind.

\subsection{Trust}

The concept of trust is a bit of a moving target with no universally agreed upon definition~\cite{huddleston2015trust}.  Different domains define trust in different ways depending on the discipline, context, or application of the concept.  Roughly speaking, we can find trust explained as a willingness to rely on another person or entity~\cite{moorman1993factors}.  Some take this a bit further, suggesting that trust has a positive connotation as it relates to confidence in a person or entity~\cite{lewicki2013role}.  Many disciplines' approaches to trust mention confidence and reliability as important factors integral to understanding trust, with some viewing trust as a choice, or rather a decision to place one's confidence in others~\cite{li2003trust}.  

When approaching trust through the lens of artificial intelligence, machine learning, and foundation models, trust becomes split across the varied components of AI systems.  With this split, definitions of trust also fracture and are defined and refined by the entities on which trust is questioned.  For instance, reliability may be an important factor in trusting the output of a foundation model irrespective of the input and model designer, whereas confidence in the credibility of the input data is likely a more accurate definition of trust when focused explicitly on training data sets.  It is important to also highlight the downstream concepts of which trust is a key component.  \textit{Credibility} is often defined as a combination of trust and expertise~\cite{hovland1951influence} which is in turn fundamental for understanding communication and persuasion~\cite{hovland1953communication}. However, definitions of trust and credibility have been forced to change as in the 1950s it was unlikely that many communication researchers were thinking about the role of credibility and trustworthiness as it relates to foundation models, let alone foundation models situated within the geographic domain.

Though not explicitly related to geographic foundation models, existing work has explored trust, via credibility, in geographic data and geographic information science more broadly.  Much of this has focused on user-generated geographic content, often referred to as volunteered geographic information (VGI).  Examining trust in this context is important as the vast majority of data on which geo-foundation models are trained is generated by \textit{users}.  Interestingly, much of the work exploring trust and credibility in user-generated geographic content stipulates that users of such data often question the quality, veracity, and overall value of such data~\cite{flanagin2008credibility,haklay2012citizen} with further work identifying the reputation of contributors as a fundamental factor in the perceived trustworthiness of the data~\cite{bishr2010can,d2014vgi,fogliaroni2018data}.

Within the broader artificial intelligence community, trust is often approached in two, very different ways.  To many in the computational sciences, trust is viewed as a transaction and is squarely situated within the security domain.  In this instance, trust is something that can be verified, often cryptographically~\cite{walton2006cryptography} and involves various forms of authentication and security protocols~\cite{shin2010effects}. While this approach to trust is very much needed for the security and credibility of systems, it is a tangential discussion to this chapter's focus on trust in foundation models within the context of geography.

The second manifestation of trust often discussed within AI is trust in the process itself.  This approach really situates trust within the broader discussion of ethics in AI~\cite{dubber2020oxford,khan2022ethics}.  In this way trust is a critical aspect of AI centered within the responsible development, deployment, and use of AI systems including the development and use of foundation models which are increasingly used as the basis for decision making. This notion of trust also closely aligns with reproducibility and replicability \cite{kedron2021reproducibility} as well as with broader discussions of fairness, bias, and algorithmic neutrality \cite{ntoutsi2020bias,janowicz20233}. This is the approach to understanding trust that we will continue to examine throughout this chapter, specifically as we explore the various \textit{types} of trust that operate in the development of foundation models.


\section{Types of trust in foundation models}

Given the wide range of definitions for trust it should not be surprising that there are various \textit{types} of trust.  Different domains have developed taxonomies or categorizations of trust~\cite{marti2006taxonomy,josang2006exploring,syropoulos2023we} that are specific to their domains.  As foundation models have emerged as a leading approach to knowledge synthesis, the concept of trust has morphed to meet this new technology.  
Prior to introducing the types, it is important to first specify the lens through which trust is explored.  Specifically, trust must always have at least two actors, the \textit{truster} and the \textit{trustee}~\cite{marsh2005trust,cho2015survey}.  The truster in most cases is the individual (e.g., you, the reader) or the entity that is relying on the knowledge produced by some process.  The trustee ranges considerable depending on the type of trust, the process, and the entity conveying knowledge.  In this chapter we focus on what we see as the three key types of trust related to foundation models and situate this trust within a geographic context.  In our three types we assume the truster as the non-technical (though likely geographically-literate) end-user of a foundation model whereas the trustee is split into the different components of foundation modeling, as shown in Figure~1. 

\begin{figure}[htb]
    \centering
    \includegraphics[width=0.7\linewidth]{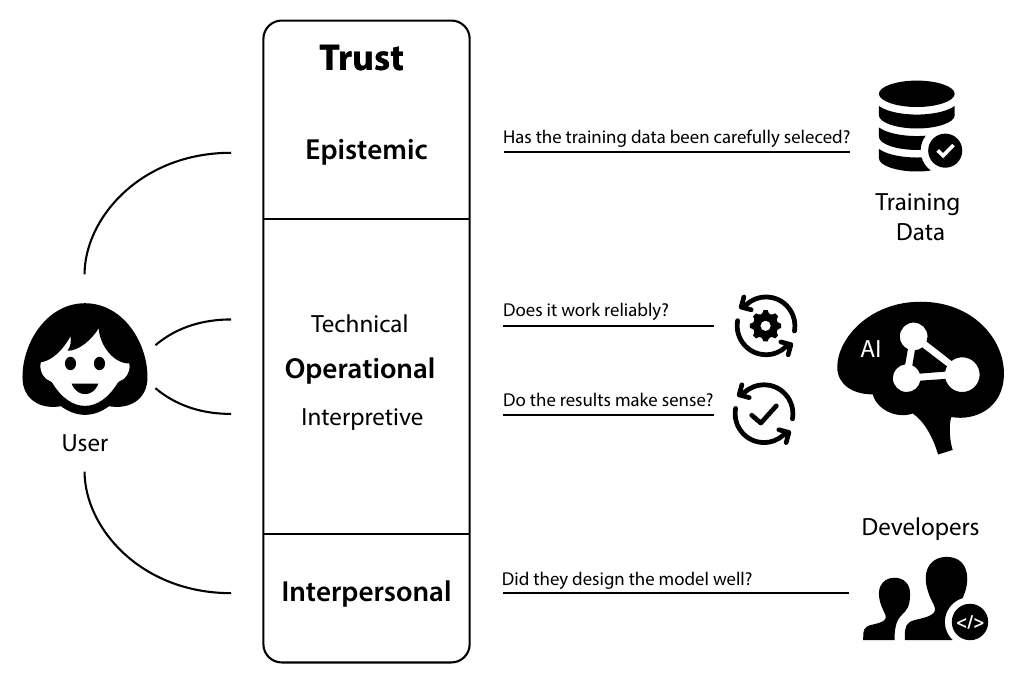}
    \caption{Types of trust and aspects of the AI model to which they are related.}
    \label{fig:enter-label}
\end{figure}

\subsection{Epistemic trust in the training data}

	This form of trust focuses on the data, where the trustee in our truster-trustee relationship is the data itself. As we know, foundation models are only as good as the input data on which they are trained~\cite{bommasani2022opportunitiesrisksfoundationmodels}, which understandably places data at the center of any discussion related to trust in a model (and the AI system potentially running on top of it).  When discussing data, it is essential to mention the various factors of bias~\cite{ntoutsi2020bias}.  Acknowledgment, discussion, and publication of bias in any aspect of foundation modeling is essential to building trust, and bias in data is the first place to start.
    
	Much of the data ingested for training models in the geospatial sciences is observation data (e.g., surface temperatures, water salinity, traffic, crime spots, etc.).  Though bias may appear to be less of a discussion in measured physical systems than human observation, it is never-the-less present.  Measurement bias arises when the tools used to collect data are either flawed or not appropriate for the task, resulting in inaccurate or distorted data.  Similarly, scaling methods such as aggregation or interpolation change a data set and bias is based on the method used or the decisions of weights determined by the modeler.  Examples of this include sampling resolution of nitrogen in soils.  For instance, a research team might collect soil samples randomly within a given district boundary and then use spatial Kriging to down-sample the resolution of the data to a regular 5-meter grid.  While interpolation methods like Kriging are common practice in the geospatial sciences, it must be acknowledge that the accuracy of the data decreases as distance from the original sample locations increases.  While this form of bias\footnote{There is, of course, substantially more to be said here about measurement error, sampling strategies to mitigate bias, observation procedures, and so on.} can be reported by the team doing the analysis, the bias and uncertainty of the data are not explicitly accounted for when used as input for a machine learning model~\cite{wang2021show}. Finally, social and cultural aspect may also lead to bias in observation data, e.g., when sampling air quality more densely in affluent communities. Similarly, measuring biodiversity is not free from (coverage) bias as most of our observations are collected in certain regions of the world and near transportation infrastructure \cite{chapman2024biodiversity}. 

	Bias is really at the societal forefront when discussing social data such as natural language text, reviews, pricing, and so on.  These sorts of data are subject to a variety of biases including selection, exclusion, and cultural biases.  Since their inception, foundation models, and most early machine learning models, have been accused of perpetuating stereotypes and cultural prejudice~\cite{lee2018detecting, GoogleCh19:online}.  The argument continues to be that selection bias occurs when the data used for training a foundation model is not representative of the full population or even the specific case.  This leads to bias and inaccuracies in the output of a model.  For instance, the vast majority of freely available data on the internet, and the basis for most foundation models, is communicated in the English language~\cite{pimienta2009twelve}, even dominating national languages for much of the world \cite{ballatore2017digital}.  Further more, most written text ingested by today's leading foundation models was generated by white males~\cite{navigli2023biases,kotek2023gender}.  Global gazetteers show significant imbalances regarding the completeness and accuracy of places between countries, with a strong bias towards the Western World~\cite{ACHESON2017309,liu2022geoparsing}. All of this potentially implies that these models are much more accurate when modeling white, english-speaking males than any other population.  This is deeply rooted in geography as socio-economic and demographic identity are squarely situated within geographic space meaning that the biases in the data used for training today's leading foundation models are biased by geography. Hence, a lot of effort is being invested to debias these models. From the perspective of trust, this leads to the question of who is debiasing and whether this process is also governed by cultural and political pressure.

    Privacy also plays an important role in bias of training data as privacy concerns related to an individual or group's data is often reported as the reasoning for not sharing or publishing information.  For instance, The majority of trajectory-based analysis currently being done by spatial data scientists today relies on publicly available datasets which tend to be those made public because they are based on ``shared'' vehicles and rely on tax-payer funding or subsidies. For instance, municipal bike sharing data (e.g., Montreal's BIXI) or taxi-cab data (e.g., NYC's taxicab and limousine commission dataset) form the basis for countless transportation research studies.  Unfortunately, few personal vehicle based trajectory datasets exist as publicly sharing personal trajectory data presents a significant privacy risk~\cite{terrovitis2008privacy}.  We have also recently witnessed ride-sharing companies like Uber remove public access to their mobility and trajectory datasets with some fearing these datasets were actively being used to train foundation models and others arguing that the privacy of their riders were being violated. All of this contributes to training data bias.  When individuals and companies are unwilling to publish their data for fear of privacy violations, the representativeness of a foundational model is put into question, as there are large gaps in the data being used to train the models.  How can broad population mobility be modeled when only shared vehicle data are used for training?

\subsection{Operational trust in the model}

    Outside of the training data used in a model, there is a second type of trust in which a truster assesses the trustworthiness of the model itself (the trustee).  This includes two sub components, namely the \textit{technical trust} and the \textit{interpretive trust}. 

    Technical trust is trust in the underlying technology used to run a foundation model.  Do I trust the model's (or the system on top) ability to perform its assigned tasks accurately and reliably based on its design and architecture of the system?  This form of trust is not unique to foundation models but is more generally a type of trust one has, or does not have, in computational systems.  This trust (or mistrust) has been amplified, however, by what is often refereed to as the \textit{black-box} of deep learning~\cite{castelvecchi2016can}.  There is a general concern among end users that as these already complex models are become even more complex, we are less inclined to know if something goes wrong, or when it goes wrong, how it went wrong.  Most of the decision-making information infrastructure that governs our society are distributed across hundreds of computer systems which are in-turn distributed geographically around the world.  The shear size of these systems requires a level of trust in technology that has not been experienced previously. It certainly makes society more fragile, especially as despite their geographic distribution most of these systems belong to a handful of companies. Aside from the inner workings, there are issues of trust related to the security of the model and data, frequency of maintenance, and reliability of the technical components.

    Users of these models (and even those not actively interacting with such models) also must trust that these models will use their private data responsibly.  There are two aspects to this type of technical trust, one related to the data that is collected about users without their knowledge, and the other related to the information that is actively \textit{contributed} to the model by the user (by way of providing context for an analysis or prompt).  We have discussed the first case with regards to privacy of the data but also must make clear that it is very rarely made transparent how much of the training data are collected.  The second is that foundation models require contextual information in order to conduct relevant analysis and provide personalized solutions to problems.  In order to do this, users of these models often provide personal information in order to received personalized, contextually-relevant results.  Aside from being used to answer a specific question, these data are ingested into the model as training data.  Users must trust that these data are being safeguarded by the companies or agencies producing the models.  Unfortunately, numerous instances have been identified where an individual's data has been made public, either through the scraping of public (or private) data sources~\cite{nytimesleak}, or through publishing context or previous prompts~\cite{techradar}.

    Interpretive trust, as the second sub component of operational trust is trust in the models ability to identify patterns in the input training data, focus \textit{attention} on the salient features of the data, and produce reasonable results that represent the underlying knowledge base.  As foundation models become more and more complex, an understanding of exactly how this occurs is becoming increasingly intransparent to the average end-user.  As the user loses their ability to \textit{follow} what is happening within such a technology, they are forced to replace their lack of knowledge with trust in the system and in the people constructing them (the topic of the next type of trust).  How much can the average person actually trust a system that ingests an incomprehensible amount of data, performs billions of operations, and outputs a recommendation or prediction?  The `magic' here lies in how modern AI systems learn from the patterns they identifies in the data.  For instance a foundation model trained or fine-tuned exclusively on a social data set from USA would extract patterns unique to the culture and use these patterns as the basis for any future analysis.  If this model were then asked to make a recommendation for government data privacy regulations in Germany, it would likely produce recommendations not consistent with the cultural norms ~\cite{zhang2023rehumanize}. This is even more important for generative AI (GenAI) where the generated output does not stem directly from any primary source. The end-users of a foundation model must trust that the data, patterns, and modeling reflect their cultural perspectives, for instance.  Unfortunately, the black-box nature of many commercial foundation models limits one's ability to audit the model, and, therefore, the user is forced to trust purely in the interpretation of the model. This also explains the recent push for more explainable (Geo)AI \cite{xing2023challenges}.


\subsection{Interpersonal trust in the modeler}

Finally, there is the trust between the trustor and the person or team that designed and implemented the foundation model, or what we here call \textit{interpersonal trust}.  While the previously mentioned operational trust is scoped to the technical and interpretive aspects of the model itself, this type of trust involves that actual decision making process that goes into developing the technologies on which foundation models are built as well as development of the models themselves.  A modeler is responsible for choosing the input data, the form of the data, determining the number of model parameters, training objectives, and so on.  As we often use the analogy of foundation models ``learning,'' we must by extension include some notion of a teacher or parent that exposes the model to new inputs and places bounds on how the model learns. In case of RAG-based systems, this also means that the selection of sources used for (factual) lookup would have to be disclosed as well. In a very noteworthy decision, OpenAI recently released its model specs to show how it addresses some of these issues.\footnote{https://model-spec.openai.com/2025-04-11.html}

There is a clear ethical aspect to interpersonal trust. A foundation model end-user must trust the selection of input data and that the decisions made by the modeler, aligns with their own ethics.  Take for instance the example of self driving cars and the morality of who should die in a car crash~\cite{Driverle80:online}.  Should a self-driving vehicle be in a situation where there is no option but to crash into, and likely kill, an elderly woman or a young boy, which should the car choose?  These are the questions being asked by ethics in machine learning researchers around the globe~\cite{awad2018moral}.  As the output of a foundation model is almost entirely built on the input data created by people, it would be important to ensure that the data mirror societal ethics related to topics such as mortality.  Whose ethics though \cite{janowicz20233}?  In most cases, foundation models rely on the ethics of the modelers themselves or a subset of like-minded individuals~\cite{lo2020ethical}. However, we know that these modelers are themselves a very unrepresentative part of society at large. This strongly suggests that the end-user needs to be both aware of the ethical leanings of the modeling team and be in agreement with those ethics.  In reality, most end-users are not made aware of such leanings, and, therefore, they must trust that the modelers are bound by some sort of ethical principles~\cite{jobin2019global}.  One can reasonably make the argument here that these decisions are cultural in origin and therefore tied to geography.  Some cultures operate on a cast-based approaches to human worth~\cite{dumont1980homo} where others approach value and worth through meritocracy.
    
While it may be easy to shrug off these concerns over trust in the modeling team as scare tactic which have a negligible impact on the results of a foundation model, take for instance the current war in the Ukraine.  At time of writing both Ukrainian and Russian software developers are developing algorithms to run on Autonomous Aerial Vehicles (AAVs).  As network jamming is often used in countermeasures, many of the software packages loaded onto these drones could soon be developed in such a way where the AAV itself is responsible for identifying a target and deciding whether or not to launch a projectile~\cite{InUkrain4:online}.  This means that modelers are responsible for building a model that has a basic understanding of the environment, the context, combatants and non-combatants, as well as the ability to kill, all without human supervision~\cite{HowAIisc64:online}.  In these cases, not only does one require trust in the technology, but the military, the government, and the general public must all trust in the ethics of the developer designing and implementing such a system. Given the scarcity of lethal resources, the decision of which targets are worth hitting (first) is not primarily an AI decision; it is a human value judgement.

Aside from these three key types of trust, there are numerous other instances where trust fits into the discussion on foundation models and GenAI.  Trust related to social norms, while similar to interpersonal trust is centered more on trusting foundational models to be developed and operate in ways that align with social norms and foster positive social outcomes. Institutional trust assumes that models comply with legal, regulatory, and institutional frameworks.  These are particularly salient in domains like healthcare and finance.  Finally, there is trust in what users will do with the predictions, recommendations, or generative outputs from foundational models.  This is a social and interpersonal trust in which the trusters are also the trustees.  In this perspective we acknowledge that we are all consumers of foundational models and that we must trust one another to use such models morally and in a way that does not negatively impact our communities. A counterexample would be the generation of false (information) videos and pictures for social media use.
 
        
        
        

\section{Importance of trust in AI and foundation models}
In this section, we provide a concise overview of the key reasons why trust is essential for foundational models. We then proceed to present a series of examples and arguments drawn from the geographic domain to illustrate these points.

\subsection{Accuracy and reliability}

Soon foundation models may serve as the backbone for decision-making processes in various geographical applications, such as urban planning, navigation, disaster response, and environmental monitoring. As in more traditional geographic information science \& systems, the ability of the analyst/researcher to rely on these models is important to ensure informed decisions.  Inaccuracies in a model's output can lead to problematic (sometimes catastrophic) consequences such as ill-informed disaster response strategies or flawed transport infrastructure planning.

Trust in the \textit{accuracy} of these models is essential, yet due to their opacity, many domain scientists are at a loss. The primary challenge here is that while domain experts may have a strong grasp of the data fed into these models, they often lack visibility into the intricacies of how the models were trained let alone how they get to their final results. This ``black box'' nature makes it difficult for implementers to fully trust the outputs. They are left relying on the assumption that the model is identifying and focusing on the most salient features of the data, even when they cannot validate the processes themselves.

The ability of the model to produce accurate results \textit{consistently} and across different scenarios is another key aspect of this trust. For example, consider a model that predicts crop yields under different fertilizer treatments. If this model is unreliable, its results become meaningless and even dangerous, given the implications for global food security and agricultural planning. In such cases, trust in the model's ability to deliver consistent, reliable outcomes is just as important as its accuracy.  Here, we can perform reproducible experiments as long as we can expect that the past is a strong indicator of the future, but what about highly dynamic systems such as (policy making in) smart cities?  This problem is further exacerbated when considering that many of the applications of foundational models are constantly improved, thereby rendering reproducibility futile even given the exact same input data.

\subsection{Transparency and explainability}

A number of issues require elucidation in this context, including transparency of how the actual models \textit{learn} but also the transparency of how the data are collected and outputs produced.  In order to develop trust, transparency must exist throughout the entire process.  There are efforts currently in the works to assess and compare various commercial foundation model frameworks using indices~\cite{bommasani2023foundation,larsson2020transparency}.  These approaches have often involved developing a range of indicators that codify transparency in foundation models including on all aspects ranging from data and labor to the details about the model itself such as size and capabilities. Further work examines the transparency of distribution methods, usage policies, and affected geographic regions~\cite{shi2023thinking}.

Recently, within AI-focused research we have seen a subdividing of the domain with some research teams aiming their efforts at eXplainability of models or what is often termed, XAI.  This thread of research is typically led with the objective of increasing trust in the process and output of models, even if it comes at the cost of accuracy~\cite{papenmeier2022s}.  Researchers in this domain posit that confidence and trust in a model is driven by a user's ability to understand the process through which an output is produced.  They suggest that more explainable models are essential for the future success of AI and by extension, foundation models~\cite{zhang2020effect,li2024geoai}.  In geography, researchers and practictioners often deal with spatially heterogeneous data (e.g., land cover, population distributions) and the results of analysis of these data have far-reaching implications for environmental management and policy-making. When the model process is interpretable, it permits researchers to audit and validate the process to ensure results that are consistent with established domain knowledge. 

\subsection{Ethical considerations}

As part of the decision making process, foundation models are increasingly used in socially and culturally sensitive contexts.  For instance shifts in human migration patterns~\cite{xue2022leveraging} or revised land-use planning~\cite{zhang2024urban} increasingly rely on decision support systems built on top of foundation models. A lack of trust in these models can disrupt the entire decision making structure as well as trust in the decision makers themselves.  Trust in this case means that in addition to trusting the accuracy of a model, users trust that a model employs the same ethical principles that they do.   Decision makers are often held to an even higher ethical standard than most, so by extension, so too must the models on which they rely.  For instance, without any external ``ethical intervention,'' a foundation model trained or fine-tuned purely on United States home insurance applications of the past has the potential to recommend \textit{redlining} as a reasonable option for mortgage lenders~\cite{hillier2003redlining}. 
    

Trust is a core component of ethics within artificial intelligence and related technologies.  This is due to how it fundamentally influences how users accept the technology, engage with it, and rely on its outputs. In foundation models, ethical concerns such as fairness, transparency, accountability, and bias are all intrinsically linked with trust. It ensures that models are not only effective but also reliable and equitable in the outputs that they produce.  The burgeoning field of Ethical AI~\cite{mittelstadt2019principles,jobin2019global} demands that AI systems be developed with society's diverse needs at the forefront and we find that trustworthiness is particularly significant in applications that impact public life, e.g., health, climate change, etc.  In these examples, any biases or limitations in the data and models have the potential to exacerbate existing social inequities~\cite{janowicz2022geoai}.  This has led to further exploration and the development of sub-fields such as Responsible AI~\cite{dignum2019responsible} or Fair AI~\cite{feuerriegel2020fair}.

\subsection{Security and risk management}

Increasingly applications built on foundation models are being given more and more responsibility as well as authority over every day systems. There are a wide range of potential issues related to granting such authority to these systems including, as we have mentioned previously, a lack of transparency in how the models are pre-trained and fine-tuning.  The centralized nature of many systems can be problematic and security of the feedback mechanisms require careful auditing in order to assuage users~\cite{rao2023building}. Take, for instance, the fact that most navigation systems today are built on custom software unique for purpose, but as foundation models improve, these specialized models will be replaced by simpler interfaces sitting on top of foundation model-based routing recommendation systems. These systems will continue to ingest massive amounts of context-rich, real-time data (e.g., traffic, weather, etc.), synthesize the relevant patterns, and calculate optimal routing across a variety of transport models.  By handing over much of the ``heavy lifting'' to these systems we are forced to put an immense amount of trust in them from a safety and security perspective.  Abstraction of the underlying process from the everyday user is a reality of advances in technologies.  A user trades knowledge of how a system works for advanced functionality.  Think for example about the benefit most users get from a mobile device without understanding the underlying technology.  Foundation models are taking a leap forward in abstracting their inner workings from even the most informed user.  

While there is a clear risk to handing over personal navigation to a foundation model, it pales in comparison to many of the other tasks that foundation model-based AI systems have the potential to execute.  The ability of AI models to handle national defense, medical diagnoses, economic forecasting, and many more are already being tested~\cite{moor2023foundation,oniani2023adopting}.  We need not look much further than AI's involvement in misinformation~\cite{monteith2024artificial, aimeur2023fake} to understand the risks involved.

\subsection{Data quality and bias}

With respect to data quality and bias, trust is especially important; foundation models often draw from large heterogeneous datasets that may not accurately represent specific subject matter due to a lack of balanced perspectives and demonstrate inconsistencies in data quality \cite{liu2024measuring}. Geographic data, in particular, can vary widely in accuracy, coverage, and representation, with regions differing in the availability and precision of data collected. For example, rural and remote areas often suffer from data scarcity, while urban areas may have dense, frequently updated data sources – particularly where the data sources rely on volunteer work. Similarly, there is plenty of existing work that has shown bias in the representation of data creators and contributors~\cite{jiang2020identifying}. This imbalance can lead to a model that performs well in data-rich areas and domains but produces less reliable outputs in data-poor regions, creating disparities that erode trust among users in underrepresented communities and groups \cite{liu2024measuring}.

In geography, biases in data can also reflect socio-political factors, such as censored or restricted data from certain governments, incomplete historical records, or data sets that under-represent marginalized communities. These biases shape how foundation models interpret spatial patterns and predict trends, potentially reinforcing existing inequalities. Trust in these models hinges on the model developers' ability to identify, disclose, and mitigate these biases. For instance, a large body of literature has investigated the biases related to predictive policing~\cite{alikhademi2022review, meijer2019predictive} with many of the studies reporting that bias in the input data often leads to biased and flawed predictions.  Transparency and efforts to mitigate bias within the training data are discussed as one possible solution to this issue. By addressing data limitations and acknowledging regional and community disparities, practitioners can create models that users find credible particularly in high-stakes applications like security, environmental risk management, and public health.


\subsection{User adoption}

Users need to trust technology in order to use it.   While this statement glosses over the complexity of users decisions to trust a technology, it is generally true.  Although the vast majority of users do not need to understand the inner workings of the model, or even the bias of the input training data, they do need to trust the output produced by a model or the actions taken as a result of those outputs.  The commercial vendors of large language models, for instance, are aware that user adoption is heavily tied to trust in the outputs of the models.  Users want to see that model outputs generally align with their world view and that when given a simple task, produce results that are verifiable~\cite{choudhury2023investigating}. Models are forced to thread the balance between generating confidence in the results and communicating uncertainty.  

Advances in AI and foundation model-related technologies is almost entirely depending on continued user adoption.  Training and re-training such models is costly, requiring financial investment from industry and government.  These sources of funds are tied to a strong and robust user-base. In fact, it is rumored that Microsoft and OpenAI agreed on a definition of Artificial General Intelligence (AGI) that is mostly based on revenue, namely that OpenAI will have achieved AGI when its AI systems generate at least \$100 billion in profits.\footnote{https://techcrunch.com/2024/12/26/microsoft-and-openai-have-a-financial-definition-of-agi-report/}  In addition, when users engage with these models, they provide feedback, insights, and data that help refine and improve the models, making them more adaptable and relevant across diverse contexts.  This engagement is essential for continual advancement.  Furthermore, user adoption arguably increases equitable access to AI technologies and the array of societal benefits that they promise.  As more users adopt the technology, there are more, diverse voices in the discussion, technology improves, and barriers to entry are eliminated. From a perspective of trust and given the increasing accuracy and ability to recall billions of facts, humans may soon trust AI more than they trust each other.

\section{A geographic outlook}

A geographic perspective on foundation models is an aspect that should be explicitly addressed.  As those who work with geographic data are aware, there are unique challenges and opportunities that spatial data presents. Geographic researchers operates on the principle that ``spatial is special,'' highlighting the importance of proximity and spatial relationships in understanding complexities in the world.  We often wheel out the ``First Law of Geography'' when discussing data analysis with colleagues from different domains.  While a central tenant of geographic analysis, we must also suggest that it is an essential part of trust-building. Users of complex models such as foundation models are limited in what they can use to build trust in a system as much of the inner working as well as the input data are abstracted from them.  What makes sense, however is that a user is more likely to trust a local model, one that reflects their own spatial realities and immediate surroundings, more than generalized, global model, as the latter may overlook critical regional differences. This leads to very interesting follow-up questions that have received very little attention so far, namely whether we all share a common experience when communicating with generative AI. To give a simplified example, systems like ChatGPT (here 4o) will (almost) always spit out Paris when asked to `Name a city'. In conversation with humans, we would likely get 10 different answers when asking 10 different people. This can contribute to a \textit{Matthew effect} where tourists may all get the very same list of suggestions when asking `Which places should I visit in Vienna', thereby further cementing the status quo. Furthermore, biases in geographic data are unique; data collected from different regions and under various political regimes may carry social and political biases that affect how well models serve communities. This then reignites the discussion on digital sovereignty, namely that regions (e.g., governments) have the rights to control the collection, storage, processing, and use of their citizens' data as well as the right to develop and deploy technologies independent of foreign influence.  All of this makes it essential for researchers to consider geographic and cultural contexts when developing and deploying foundation models. Despite this, there is hardly any work discussing the geographic aspects of AI alignment.

Researchers in geography and related disciplines have long studied trust, credibility, and expertise outside of AI and foundation models~\cite{murphy2006building, greenberg1999geographical}, especially through the evolution of local knowledge, community-driven data, and the role of VGI~\cite{kessler2013trust,fast2024giscience}. Unlike traditional, authority-based models of knowledge (where trust rested with institutional data sources), VGI and user-generated content challenge conventional notions of authority, leading us to rethink whose data and expertise are \textit{trustworthy}.  In addressing geographic data bias in foundation models specifically, researchers should continue to explore transparency and explainability approaches that permit users to understand the sources of uncertainty and bias.  A lot of this can be done by leveraging geography itself.  General properties of geographic or geotagged data such as spatial autocorrelation can be used to validate outputs of a model thus increasing trust in the process.  Agreements between models built on diverse and representative data sets from different geographical regions can be used to mitigate unintended bias through data augmentation. User feedback and community validation processes are also critical for adapting models to reflect local knowledge, especially for community-driven projects that necessitate regional accuracy and trustworthiness.

As emerging technologies, like Retrieval-Augmented Generation (RAG), reshape the field, there is plenty of potential to enhance trust in foundation models by integrating real-time, local, context-specific data into outputs via future GeoRAG systems (e.g., developed on top of knowledge graphs).  The acquisition and inclusion of these data require an understanding of the regional importance of different data sources and the geographic-context of the problem.  These technologies also bring challenges in terms of model accountability and data provenance, prompting a need for policies that govern their ethical use.  Citizens of different regions may have varying levels of trust in government data, and this mistrust can extend to AI models that draw on these sources. Moving forward, geographers and those in related disciplines have an opportunity to help develop regionally-informed policies that prioritize transparency and respect the privacy of individuals and groups.  In building trust, we also have the opportunity to leverage our domains' traditional expertise in community engagement and citizen science, ensuring that AI and foundation models serves a diverse user base while upholding the trust necessary for equitable geographic applications.

%
%
%
%
%
%
%
%
%
%
%
%

\subsection*{Recommendations}

Based on the arguments above, we believe that the following aspects deserve special attention from our community as we learn how generative AI systems represent and reasons about geographic space and how this will shape the interaction of future autonomous GIS analysts/agents (\textit{GeoMachina}, \cite{janowicz2020geoai}). While these recommendations can, and should, be realized in a number of ways, regulation of foundation models and other AI technologies either through government or non-governmental organizations clearly needs to be examined further. We can make all the recommendations that we want, but regulation and oversight is necessary to ensure that (Geo)AI provides equitable benefits for our societies.  Regulation ensures a level of trust in a technology or project~\cite{six2013trust} as users can feel confident that oversight exists.

First, \textbf{transparency is a priority}. In order for trust to be established or maintained between users and future generations of foundation models and AI, models must be more transparent.  This includes transparency at all steps in the process from data collection and parameter selection to the modeling process itself. For geographers this also relates to transparency in the data collection process and reporting regional variations in data. We acknowledge that there is an ongoing debate as to how important transparency is in relation to other factors (e.g., accuracy, reproducibility), but purely from a trust perspective, we argue that transparency and interpretability are paramount.

Second, \textbf{bias needs to be acknowledged and  mitigated}.  Bias is an inherent part of any domain; while often seen as negative, it can sometimes provide valuable insights into specific aspects of a problem.  From a geographic perspective, we know that geographic bias is important for helping models fit local social norms, legal requirements, etc. However, reducing representation bias in data is important when striving for equitable outcomes in model predictions. As we continue to make advances in GeoAI, it is essential that we openly acknowledge bias in all aspects of foundation models and continue to work to mitigate it where appropriate. We must also ask `who are the engineers debiasing models, why, and how'.

Third, \textbf{uncertainty needs to be communicated}. Uncertainty exists in all aspects of model development ranging from data quality and model assumptions to algorithmic complexity.  Uncertainty also compounds as these facets interact. Clear communication about these uncertainties, e.g., visualizations, will help end users understand the confidence level of a model's outputs and contextualizes the results.  Geographers and cartographers have a long history of examining and reporting uncertainty in geographic data and analysis.  This expertise should be leveraged in developing new ways to explain uncertainty in foundation models.

Finally, and potentially controversially,  \textbf{the ethical use of data and models needs to be guaranteed}.  It is essential to recognize that not all available data is suitable for training a foundation model.  Model developers have an ethical responsibility to ensure that any human-related data respects the privacy, autonomy, creativity, and consent of the individuals involved. For example, training on medical records without patients' explicit consent clearly violates their privacy. There are other cases where this is not as clear and the race to improve model accuracy can sometimes lead to privacy infringements, where individuals' rights are compromised. As geographers we know that spatial autocorrelation can lead to reduction in privacy or individuals as knowing something about my neighbor means one knows something about me.  Further examination and acknowledgment of these types of issues in data use are necessary in developing ethical models.


\bibliographystyle{vancouver}
\bibliography{trust_refs}

\end{document}